\documentclass[aps,twocolumn,10pt,floatfix]{revtex4-1}

\usepackage[utf8]{inputenc}
\usepackage{amsfonts}
\usepackage{amsmath}
\usepackage{amssymb}
\usepackage{amsthm}
\usepackage{fontenc}
\usepackage{graphicx}
\usepackage{xcolor}
\usepackage{textcomp}
\usepackage{epstopdf}
\usepackage{braket}
\usepackage{mathtools}
\usepackage{amsmath}
\usepackage{dcolumn}
\usepackage{multirow}
\usepackage{soul} 
\DeclareGraphicsExtensions{.pdf,.eps,.png,.jpg,.mps}
\usepackage[colorlinks=true, %
            linkcolor=blue,%
            urlcolor=blue,%
            citecolor=blue,%
            filecolor=blue,%
            bookmarksopen=true,%
            pdfauthor={DZ AL LR PAO},%
            pdftitle={Spin and Valley filters on two-dimensional WSe$_2$ heterostructures},%
            pdfsubject={},%
            pdfpagemode=UseOutlines]{hyperref}

\begin{document}

\title{Spin and valley filter based on two-dimensional WSe$_2$ heterostructures}
\author{D. Zambrano}

\affiliation{Departamento de F\'isica, Universidad T\'ecnica Federico Santa Mar\'ia, Casilla 110-V, Valpara\'iso, Chile.}

\author{P. A. Orellana}
\affiliation{Departamento de F\'isica, Universidad T\'ecnica Federico Santa Mar\'ia, Casilla 110-V, Valpara\'iso, Chile.}

\author{L. Rosales}
\affiliation{Departamento de F\'isica, Universidad T\'ecnica Federico Santa Mar\'ia, Casilla 110-V, Valpara\'iso, Chile.}

\author{A. Latg\'e}
\email{alatge@id.uff.br}
\affiliation{Instituto de F\'isica, Universidad Federal Fluminense, Niter\'oi-RJ, Brazil.}

\date{\today}

\begin{abstract}
In this work, we investigate spin and valley transport properties of a WSe$_2$ monolayer placed on top of a ferromagnetic insulator. We are interested in controlling the transport properties by applying external potentials to the system. To obtain spin an valley polarizations, we have considered a single and a double barrier structure with gate potentials. We have analyzed how the efficiency of these polarized transport properties depend on the gate-potential  intensities and geometrical configurations. Additionally, we investigate how the spin and valley transport properties are modified when an ac-potential is applied to the system. We have obtained a controllable modulation of the spin and valley polarizations as a function of the intensity and frequency of the ac-potential, mainly in the terahertz range. These results validate the proposal of double quantum well structures of WSe$_2$ as candidates to provide spin and valley dependent transport within an optimal geometrical parameter regime.

\end{abstract}

\maketitle

\section{Introduction}
Transition metal dichalcogenides (TMDCs) play an important role in nanotechnology nowadays due to their novel physical and chemical properties \cite{Norden2019,Qi2015,Tahir2017,Hai2014,tmdc3a,tmdc1,tmdc2,tmdc3}. Compared with graphene, they have the great advantage of presenting electronic gaps in a wide size range within the visible and infrared spectra. Examples of TMDCs are tungsten disulfide (WS$_{2}$), molybdenum diselenide (MoSe$_{2}$), tungsten diselenide (WSe$_{2}$), and molybdenum disulfide (MoS$_{2}$) \cite{tmdc12,tmdc13,tmdc14,tmdc15,tmdc17,tmdc18,tmdc19,tmdc20,tmdc21,tmdc22}. These materials have been considered for many technological applications in electronic \cite{tmdc4,tmdc5} and optoelectronic \cite{tmdc6,tmdc7,tmdc77}, as gas sensing \cite{tmdc8} devices, ultrasensitive photodetectors \cite{tmdc9}, among others  \cite{tmdc10,tmdc11,tmdc12,tmdc13,tmdc14,tmdc15}.
In particular, the TMDCs monolayer of group VI present a direct bandgap in the optical range suggesting possible applications in optoelectronics devices. In this context, TMDCs have emerged as excellent candidates of ultra-thin semiconductor materials, with a tunable bandgap in the optical region \cite{tmdc1,tmdc4}. Moreover, the application of appropriate gate voltages and the presence of magnetic materials to induce magnetic proximity effects have been explored, revealing the different mechanisms of tailoring electronic properties \cite{Aivazian2015}.

Electronic transport properties of TMDCs and their possible applications are determined by the corresponding carrier mobility and spin and valley carriers mean free path \cite{tmdc2}. There are several challenging proposals involve the possibility of spin and valley filtering within spintronic and valleytronic scenarios\cite{Yamping2019}  The theoretically predicted values for the charge mobility are promising, but, of course, limited by intrinsic scattering processes such as phonon scattering or local Coulomb potentials induced by impurities \cite{tmdc1,tmdc2,tmdc3,tmdc4,tmdc3a}.  Besides, Spin-orbit coupling induced valley Hall effects were reported on TMD materials and addressed as a manifestation of applied gate voltage\cite{Zhou2019}. This type of induced valley Hall effects was attributed to a coexistence of Ising an Rashba spin-orbit coupling in gated/polar TMDCs originated from inversion-asymmetric spin-orbit interactions.

Although most of the studies on transport properties of TMDCs systems have been made in the \textit{dc} regime, interesting quantum transport phenomena appear when external time-dependent fields perturb these materials. Comparing with stationary potentials, a time-varying one can effectively modulates the quantum phase of the electronic wave functions \cite{t1,t2,t3}, bringing new possibilities for technological devices.
Single-electron pumps were designed applying a time-dependent gate voltage in quantum systems \cite{t4} and radio frequency analog electronic devices were synthesized based on carbon nanotube (CNT) transistors \cite{t5}. Others interesting effects are the photon-assisted tunneling in graphene bilayers \cite{t6}, \textit{ac}-field effects on the conductance and noise of CNT and graphene nanoribbon devices in the Fabry Perot regime \cite{t8,t88}, quantum charge pumping in carbon-based devices \cite{t9}, and irradiated graphene as a tunable Floquet topological insulator \cite{t11}.

In TMDCs, the photon-assisted transport response has been lesser reported in comparison with other materials. For instance, the electric behavior of a mechanically exfoliated  MoS$_{2}$ single-layer based phototransistor \cite{tmt1} was investigated. The photocurrent generation was found to depend solely on the illuminating optical power at a constant drain or gate voltage, exhibiting better photoresponsivities than those proposed in the graphene-based device. Hybrid TMDCs photoelectronic devices, based on graphene-MoS$_{2}$, have been proposed\cite{tmt2} as a possible application in multifunctional photoresponsive memory devices.  Also, photocurrent was found at zero bias voltage \cite{tmt3} in $p-n$ vertical junction formed by a hybrid system of WSe$_{2}$/MoS$_{2}$, as the system is irradiated by a $514$ nm laser ($5$ $\mu$W). Studies of the influence of an optical field on spin and valley polarized transport of WSe$_2$ monolayer were recently reported taking into account also the effects of a Fermi velocity barrier\cite{Hao2020}. All these reports undoubtedly suggest that photon-assisted phenomena can be an additional mechanism of controlling transport in TMDCs nanostructures.

With the above motivations, here we investigate the possibility of inducing and controlling spin and valley polarizations on different potential profiles of 2D-WSe$_2$ heterostructures. 
We have studied the case of a single and a double potential barrier configurations. We focused on the resonant regime and how this effect allows the spin and valley polarizations. Exchange valley splitting was provided by magnetic proximity effect with the advantageous that the splitting is dictated by the exchange interaction strength and no applied magnetic field is need. This allows a convenient scenario for obtaining tunning processes. Moreover, due to the versatility of being either positive or negative valued this description opens the possibility of tunning valley splitting sign and magnitude together.
Furthermore, we have analyzed the possibility of promoting valley and spin polarization inversions on TMDCs layers under the effects of time-dependent external potentials, such as time-oscillating gate voltages or laser radiation. Different mechanisms are analyzed to synchronize the proposed system's physical parameters, such as Fermi energy, frequency and amplitude of the time-dependent potential, and external gate voltages, to optimize time-dependent transport properties, such as induced switching effects of the transport of the systems.

\section{The Model}

\begin{figure}[hb]
\centering
\includegraphics[width=1.00\linewidth]{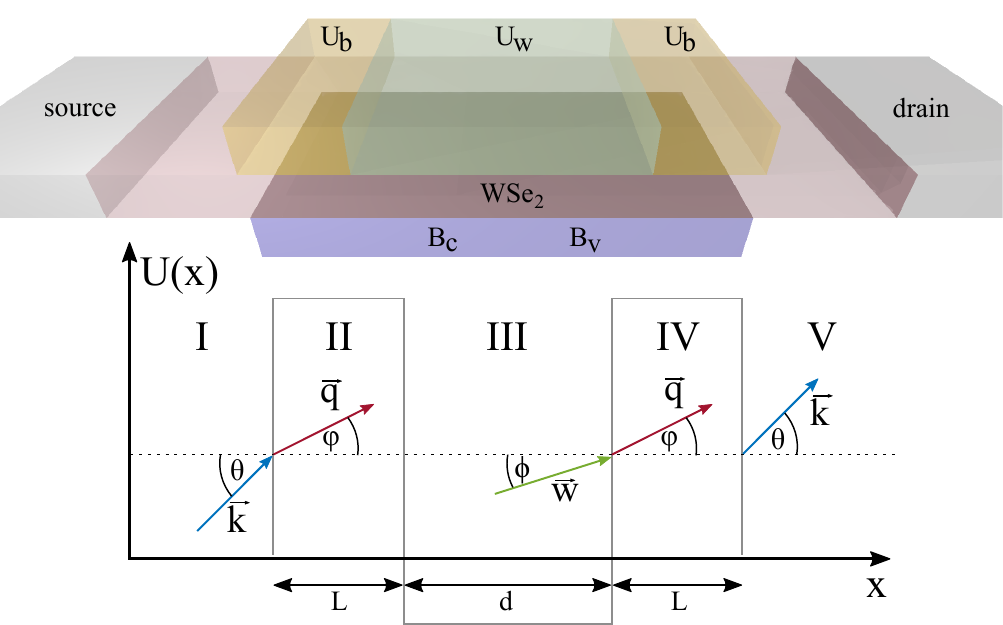}
\caption{Schematic view of a WSe$_2$ double barrier heterostructure. Zones II and IV are the potential barriers of length $L$ and height $U_b$. Region III is the well with length $d$ and depth $U_w$. In zones II, III and IV the presence of the substrate provides the exchange potentials $B_c$ and $B_v$.}
\label{fig.scheme_2b}
\end{figure}

The proposed device consists of a WSe$_2$ monolayer partially placed over a ferromagnetic insulator (EuS), as it is schematically depicted in Fig. \ref{fig.scheme_2b}. The whole system is composed of three regions: two leads, modeled by pristine regions of WSe$_2$ and a central conductor at which a ferromagnetic insulator is fixed under the WSe$_2$ layer. In order to modulate the transport response of the system, we have considered top gate potentials in different configurations, forming two distinct heterostructures: i) a single barrier and b) a double barrier. In the case of a single barrier, a top gate is placed over the finite region of the central conductor, whereas for the double barrier configuration, three different top gates are used to define this system, as it is marked in yellow/green in Fig.\ref{fig.scheme_2b}

The Hamiltonian describing the 2D system of interest can be written as:
\begin{equation}
	H = H_{0} + H_\text{ex} + H_\text{g}+ H_\text{ac}\text{,}
	\label{H}
\end{equation}
where $H_0$ is an effective Hamiltonian, written in the continuum model as discussed in previous works \cite{Norden2019,Qi2015}, which adjusts well with DFT calculations. This is a convenient approach to study transport response near the Fermi energy. The term $H_\text{ex}$ represents the exchange field induced by the interaction of the WSe$_2$ layer with the EuS ferromagnetic insulator, that promotes an energy exchange splitting as a magnetic proximity effects. The two last terms correspond to an external gate voltage $H_\text{g}$,  that can spatially modulate tunneling and resonance features and $H_\text{ac}$ is a time-dependent potential.

In the low energy approximation, the effective Hamiltonian is written as,
 \begin{equation}
 	H_{0} = v_{F} \left( \eta\sigma_x p_x + \sigma_y p_y \right) + \frac{m}{2}\sigma_z+\eta S_z \left( \lambda_c\sigma_+ + \lambda_v \sigma_- \right)\text{.}
 	\label{H0}
 \end{equation}
where $v_{F}$ is the electronic Fermi velocity and $\eta=\pm 1$ reads for $K$ and $K'$ valleys. The second term is the mass term which breaks the inversion symmetry and  $S_z=\pm 1$ is the spin index. The parameters $\lambda_{c,v}$ give the spin-splitting of the conduction and valence bands, respectively, and is due to intrinsic spin-orbit coupling, and:
\begin{equation}
	\sigma_\pm = \frac{1}{2}\left( \sigma_0 \pm \sigma_z \right)
	\text{,}
\end{equation}
with $\sigma_0$ denoting the identity matrix. The exchange term, following Norden \textit{et. al.} \cite{Norden2019} and Qi \textit{et. al.} \citep{Qi2015}, is written as:
\begin{equation}
	H_\text{ex} = - S_z \left( B_c\sigma_+ + B_v\sigma_- \right)
    \label{Hex}
	\text{.}
\end{equation}
where $B_{c,v}$ may be interpreted as an effective Zeeman field experienced by the conduction ad valence bands of the WSe$_2$ due to the proximity with the ferromagnetic substrate. It is important to mention that the Norden model parameters are based on DFT calculations to reproduce the exact band structures of WSe$_2$ in the low energy regime.

The external gate voltage term is defined as
\begin{equation}
	H_\text{g} = U(x)\,\,,
    \label{Hg}
\end{equation}
where $U(x)$ represents a sequence of potential barriers and well which define an heterostructure device, as it is illustrated schematically in Fig. \ref{fig.scheme_2b}, for the case of a double barrier system. We have denoted the barrier length as $L$ and the potential barrier height as $U_b$, whereas, the well length is taken as $d$ and the potential well deep is given by $U_w$. 
Finally, the time-dependent Hamiltonian term is given by $H_{ac} = eV(t) = V_\text{bias}\,cos(\omega t)$, where $V_\text{bias}$ and $\omega$ represent the ac-potential intensity and the frequency, respectively. It is important to mention that, following the Tien-Gordon approach \cite{tien1963multiphoton}, this term has been applied to the whole system.

\begin{figure*}
\centering
\includegraphics[width=1.0\linewidth]{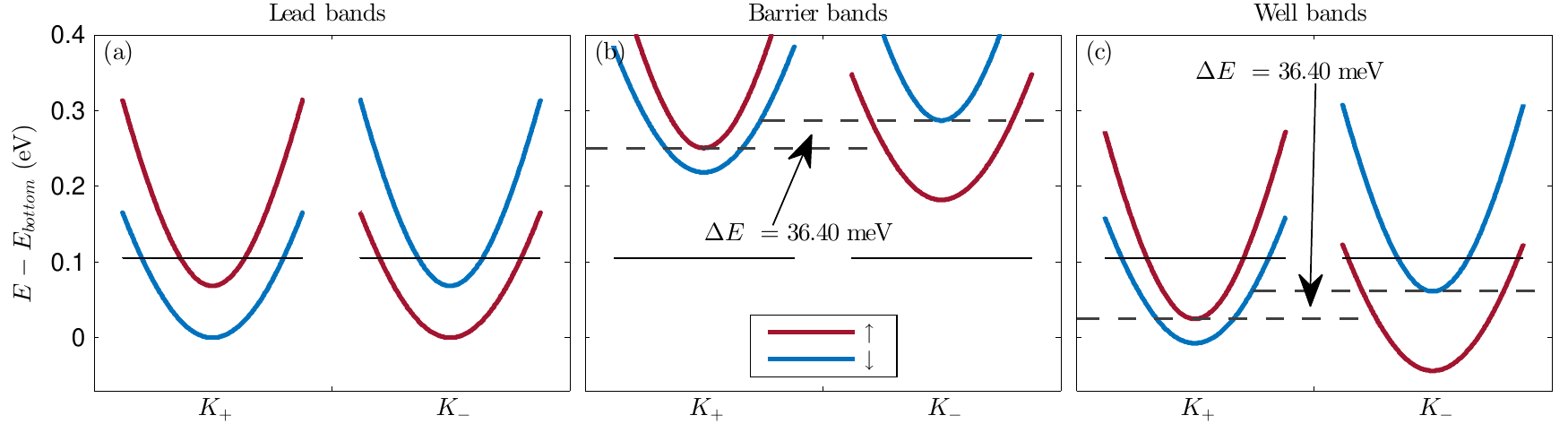}
\caption{Band structures of a WSe$_2$ heterostructure corresponding to: (a) the leads, (b) barrier region and (c) well region composing a double barrier system (in a single barrier system panel (c) is neglected). The bottom of the lead conduction band has been redefined as the zero energy. Solid black lines show the energy of the incoming electron. Dashed black lines in (b) and (c) define the energy difference between the bottom of the spin-up conduction band in the valley $K_+$ and the spin-down conduction band in the valley $K_-$.}
\label{fig.bands_WSe2}
\end{figure*}
In our model we have not included the Rashba interaction in the Hamiltonian,  considering the exchange coupling more relevant in our approach. In the calculation, we are interested in the electronic transport through the conduction band that presents an energy splitting  greater than the usual Rashba term \cite{Zhou2019}.

We have considered the cases of a single barrier and a double barrier potential profile. The electronic transmission through the conductor region is calculated following the same scheme, in both configurations, as is described in Appendix A. The transmission coefficient of the single barrier $t_{\eta,S_z}^\text{single}$, can be analytically determined, and is given by:
\begin{equation}
	t_{S_z,\eta}^\text{Single} = \frac{ 2 }{ \xi } \cos \left( \theta \right) \cos \left( \varphi\right) e^{-ik_xL}\text{,}
	\label{tsingle1}
\end{equation}

where the term $\xi$ is defined as:

\begin{eqnarray}
	\xi = &~&\cos\left( \theta-\varphi\right)e^{iq_xL} + \cos\left(\theta+\varphi\right) e^{-iq_xL} \nonumber\\\nonumber\\
	      &-&i \left( \frac{c_k b_q}{c_q b_k} + \frac{c_q b_k}{c_k b_q} \right) \sin\left(q_x L \right) \text{,}
	      \label{tsingle2}
\end{eqnarray}
with $\theta=\arctan(\frac{k_y}{k_x})$, $\varphi =\arctan(\frac{k_y}{q_x})$, $k_{x,y}$ and $q_{x,y}$ are the wave vector components of the electron outside and inside the barrier, respectively, and $c_k, c_q, b_k , b_q$ are amplitude coefficients of the electronic wave function outside and inside the barrier, defined in the Appendix A. It is also possible to obtain an analytic expression for the double barrier transmission coefficient $t_{S_z,\eta}^\text{Double}$, derived in the Appendix in Eq. (\ref{t_double}).

For both cases ($t_{S_z,\eta}^\text{Single}$ and $t_{S_z,\eta}^\text{Double}$) the angular dependent transmission probability $T_{_{S_z,\eta}}(\theta)$ is defined as:
\begin{equation}
	T_{_{S_z,\eta}}(\theta) = \left| t_{S_z,\eta} \right|^2 \,\,,
\end{equation}
while the spin- and valley- conductance are written as:
\begin{equation}
	\mathcal{G}_{_{S_z,\eta}} = \int_{-\pi/2}^{\pi/2} T_{_{S_z,\eta}}\cos(\theta)~d\theta\text{.}
	\label{Ttheta}
\end{equation}

We discuss further the effects of an ac-potential of frequency $\omega$ and intensity $V_\text{ac}$ on the transport properties of the proposed device. 
Following the Tien-Gordon formalism  \cite{tien1963multiphoton}, the spin- and valley- conductance may be written as \cite{t6}
\begin{equation}
	G_{_{S_z,\eta}} = G_0\sum_{m=-\infty}^\infty J^{2}_m\left( \frac{eV_{\text{ac}}}{\hbar\omega}\right) \mathcal{G}_{_{S_z,\eta}}(\varepsilon_f+m\hbar\omega)\text{,}
\end{equation}
where $J_m$ are the Bessel functions of first kind.

Here we use the spin dependent conductance as:
\begin{equation}
	G_{_{\uparrow(\downarrow)}}	= G{_{\uparrow(\downarrow) K}} + G{_{\uparrow(\downarrow) K'}}\text{,}
\end{equation}
and the valley dependent conductance as:
\begin{equation}
	G_{_{K(')}}	= G{_{\uparrow K(')}} + G{_{\downarrow K(')}}\text{.}
\end{equation}

Following standard definitions, the valley-polarized conductance is written as
\begin{equation}
	P_{{\nu}}	= \frac{ G{_{K}} - G{_{K'}} }{ G{_{K}} + G{_{K'}} }\text{,}
	\label{polv}
\end{equation}
whereas the spin-polarized conductance as
\begin{equation}
	P_{_{S}}	= \frac{ G{_\uparrow} - G{_\downarrow} }{ G{_\uparrow} + G{_\downarrow} }\text{.}
	\label{pols}
\end{equation} 

In what follows, we discuss the results for spin and valley polarizations obtained for both single and double barrier heterostructure systems, considering different mechanisms to modulate the transport response of the proposed device. 

\section{Results}

The low energy band structures for the  WSe$_2$ heterostructure are presented in Fig. \ref{fig.bands_WSe2} considering the three regions: leads, barriers and well. We adopt, for simplicity, the notation of $K_+$ and $K_-$, for the traditional $K$ and $K'$ valleys. Red and blue curves denote the spin up and down bands that, due to the symmetries of WSe$_2$, are inverted for $K_+$ and $K_-$ valleys.
To obtain these band structures, we have used the following parameter values: $\lambda_c = 34.20$ meV, $\lambda_v = 418.05$ meV, $B_c = 18.20$ meV, $B_v = 13.75$ meV, and $m = 1558.70$ meV \cite{Norden2019}. The lattice constant is $a_0 = 0.3316$ nm and the Fermi velocity $v_f = 5.30\times 10^{14}$ nm/s.

The energy split induced by the ferromagnetic substrate ($B_c$ and $B_v$) for the well and barrier regions, is marked with dashed black lines in the electronic band structures presented in Fig. \ref{fig.bands_WSe2} b) and c).  The energy split $\Delta E = 36.40$ meV matches the ones reported in Fig. 7 of Ref. \cite{Norden2019}. 
The energy axis has been redefined as $E-E_{bott}$ where $E_{bott}=745.15$ meV is the bottom energy of the conduction bands in the leads. The horizontal black lines in the three panels of Fig. \ref{fig.bands_WSe2} represent the incoming electron energy level, which is $104.68$ meV ($175.10$ meV) above the Fermi level for the double (single) barrier system.
Energy values up to $400$ meV were considered for the potential $U_b$ of the double barrier systems, that are chosen in a symmetric configuration. In the case of single barrier systems, $U_b$ takes values up to $200$ meV. For the well potential $U_w$ values up to $100$ meV were considered. In this context, it is possible to observe in panels b) and c) of Fig. \ref{fig.bands_WSe2} how the spin bands are displaced as the potentials $U_b$ and $U_w$ are applied in the system.
\begin{figure}[ht]
\centering
\includegraphics[width=1.0\linewidth]{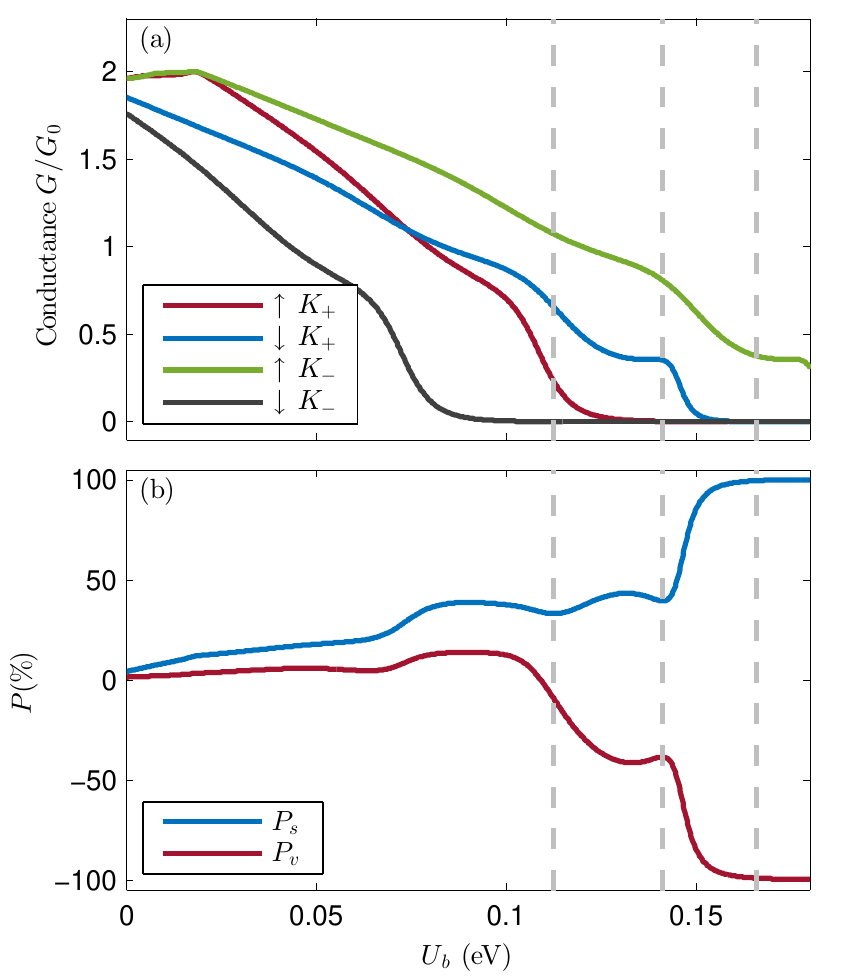}
\caption{Spin and valley conductances (a) and polarizations (b) as a function of the barrier height $U_b$ for the single barrier system, with barrier length $L = 20$ nm. Both panels are at energy $175.10$ meV above the Fermi level.}
\label{fig.1b_GP}
\end{figure}
\begin{figure}[ht]
\centering
\includegraphics[width=1.0\linewidth]{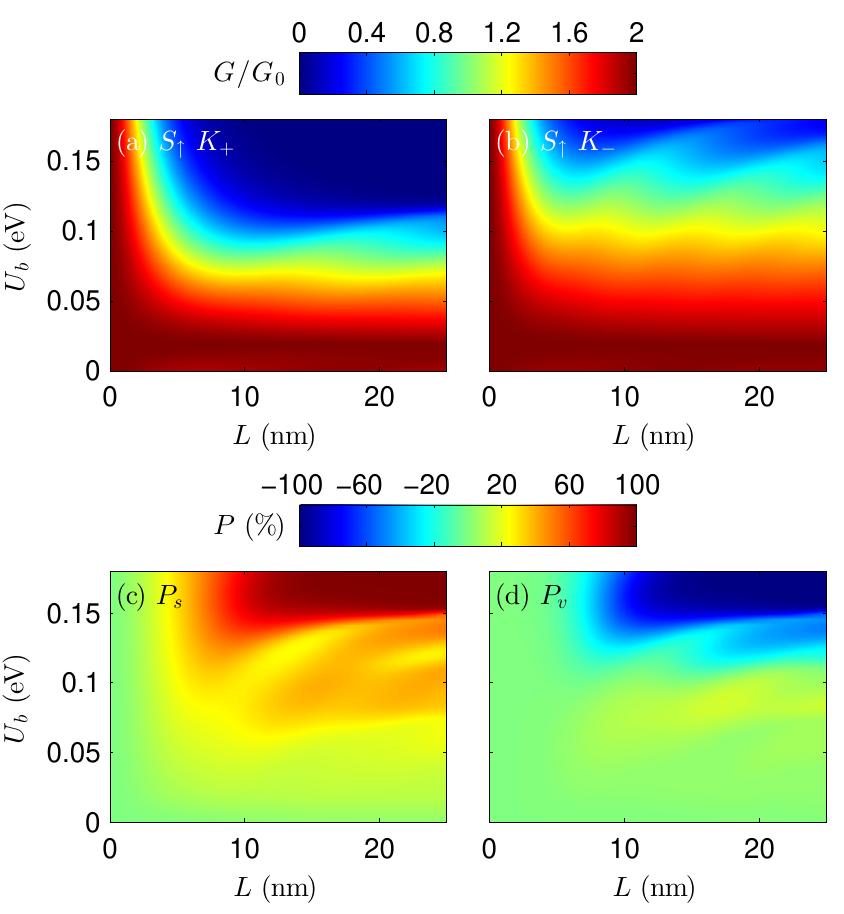}
\caption{(a) S$\uparrow K_+$ conductance component, (b) S$\uparrow K_-$ conductance component and (c) spin and (d) valley polarization contour plots, for a single barrier system, as functions of the barrier height $U_b$ and length $L$, at a fixed energy of $175.10$ meV above the Fermi level.}
\label{fig.1b_contour_GP_U_L}
\end{figure}

We discuss first the cases of single and double barrier heterostructures in the  absence of the time-dependent term, presented in Eq. 1. The ac-field contribution is only taken into account in subsection (C) where a double-barrier system is revisited under the effects of a time-dependent radiation.

\subsection{Single Barrier}

Results of spin and valley resolved conductance and polarization, as a function of the single barrier potential energy, are presented in  Fig. \ref{fig.1b_GP} (a) and (b). Here we have chosen a barrier length $L=20$ nm and an incident electron energy $E = 175.10$ meV above the Fermi level. The four spin and valley conductance components are denoted as $\uparrow K_+$,  $\downarrow K_+$, for the valley $K_+$ and as $\uparrow K_-$ and $\downarrow K_-$, for the $K_-$. All conductance curves fall down, as it is expected, as the barrier height increases blocking the electronic transport. The corresponding spin and valley polarizations, given by blue and red curves respectively, depend strongly on the barrier height, as it is shown in Fig.\ref{fig.1b_GP}(b). It is important to note that the polarization is not very pronounced when the four conduction channels are active, mainly the valley polarization. However, as the conductance starts falling down, the polarization is enhanced, achieving maximum values when there is just a single conduction channel (green curve in Fig. 3(a)). Particular potential values ($0.16< U_b< 0.20$) predict total polarized spin and valley configurations. A full spin and valley filter can be achieved then in this single barrier heterostructure.

\begin{figure}[h]
\centering
\includegraphics[width=1.0\linewidth]{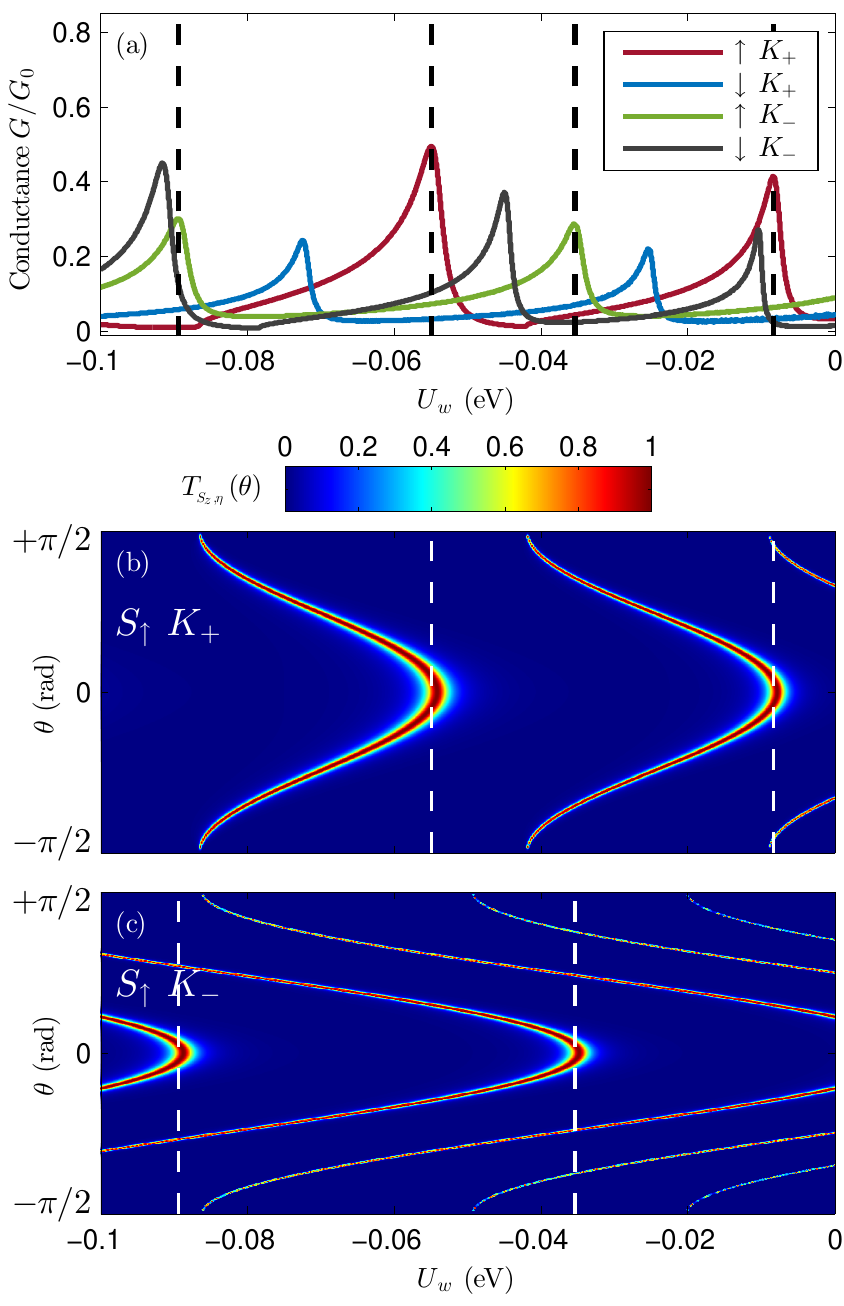}
\caption{(a) Spin and valley conductance as a function of the well depth for a double barrier system. Barrier length $L=3$ nm, well length $d=30$ nm, and barrier height $U_b=300$ meV. (b-c) Transmission as a function of the incident angle $\theta$. White dashed lines correspond to the peaks of the conductance components shown in panel (a) with black dashed lines. All panels are at $104.68$ meV above the Fermi level}
\label{fig.2b_G}
\end{figure}
We wonder about how robust this spin and valley filter are as a function of the barrier length and the range of potential intensities that we have used. In Fig. \ref{fig.1b_contour_GP_U_L} we show contour plots of: (a) $\uparrow K_+$ and (b) $\uparrow K_-$ conductance components, (c) spin polarization $P_s$ and (d) valley polarization as functions of the barrier height $U_b$ and length $L$, at a fixed energy of $175.10$ meV above the Fermi level. 
In panels (a) and (b) it is possible to observe that the $\uparrow K_+$ and $\uparrow K_-$ conductance components behave uniformly as the barrier length is increased over $5$ nm, diminishing they values as the potential intensity $U_b$ is higher. However, there is a different potential intensity at which these two components become zero, around $0.1$ eV for $\uparrow K_+$ and $0.15$ eV for $\uparrow K_-$ respectively. Considering the equations \ref{pols} and \ref{polv}, it is possible to obtain a perfect spin and valley filters in a wide range of parameters, as it is reflected in panel (c) and (d) of Fig. \ref{fig.1b_contour_GP_U_L}. Thus, single barrier heterostructures larger than $10$ nm and with potential energy range between $140-200$ meV are appropriated to obtain spin and valley filters, as it is indicated by red and blue regions in both polarization contour plots. These results give a guideline to choice  single barrier WSe$_2$ heterostructures that would behave as filter.

\begin{figure*}[ht]
\centering
\includegraphics[width=1.0\linewidth]{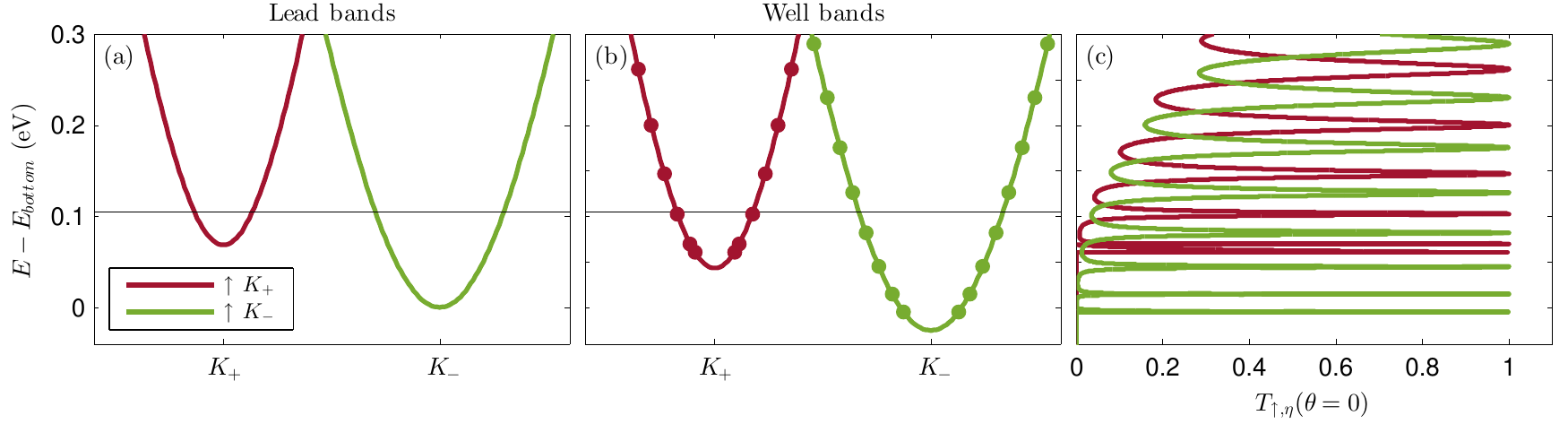}
\caption{(a-b) Lead and well bands (spin up), for the case of a double barrier system with barrier and well lengths $L=3$ nm and $d=30$ nm, respectively, barrier height $U_b=300$ meV and well depth $U_w=-25.40$ meV. The energy of the incoming electron, chosen as $104.68$ meV above the Fermi level, is marked as black horizontal lines. (c) Spin-valley transmission as a function of the incoming electron energy for the double barrier system, in the case of normal incident angle $\theta=0$.}
\label{fig.6}
\end{figure*}

\begin{figure}[ht]
\centering
\includegraphics[width=1.0\linewidth]{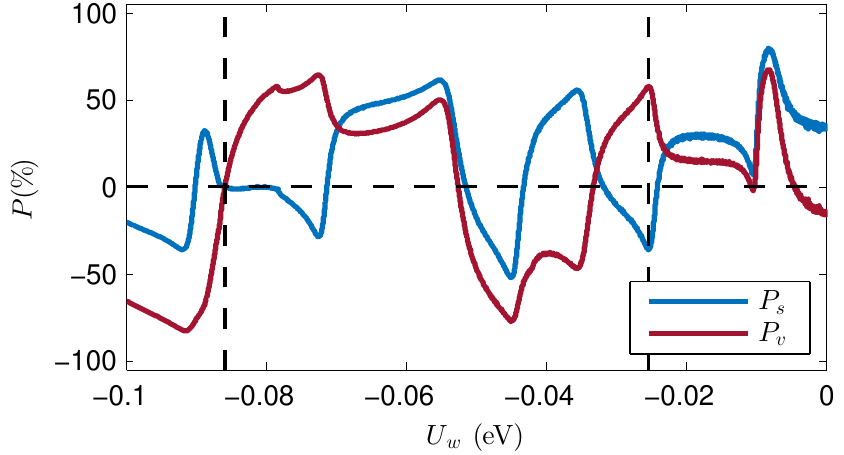}
\caption{Spin and valley polarizations as a function of the well depth for the double barrier system. Barrier length $L=3$ nm, well length $d=30$ nm, and barrier height $U_b=300$ meV at $104.68$ meV above the Fermi level. Vertical dashed lines in panel are drawn at $U_w = -85.9$ meV and at $-25.4$ meV.}
\label{fig.2b_G2}
\end{figure}

\subsection{Double Barrier}
In what follows, we explore different configurations for the top gate voltages applied to the WSe$_2$ monolayer that transform the system into semiconducting quantum well heterostructures. Results for the spin and valley conductances and polarizations as a function of the well potential energy are shown in Fig. \ref{fig.2b_G} and Fig. \ref{fig.2b_G2}  for a symmetric double barrier device. The left and right barrier lengths and potentials are equal to $L=3$ nm and $U_b=300$ meV, respectively. Although apparently small, such a size barrier corresponds to around $10$ atomic layers, which is experimentally feasible nowadays\cite{Sara2020}. The well length is fixed and equal to $d=30$ nm. Unlike the tunneling phenomena observed in the single barrier case, resonant states are typical for double barrier profiles, which define the transport behavior of this kind of resonant systems. The states may be tuned, for instance, by changing the well potential depth. This is evidenced in the sequence of conductance peaks depicted in Fig. \ref{fig.2b_G} (a). The features of the spin and valley components of the conductance are similar, although shifted in relation to the potential depth. This is understood by the alignment of the allowed conduction channels inside of the well region (defined by the well length) and the incident electron energy, as the potential $U_w$ is modified (see Fig. \ref{fig.bands_WSe2}).

It is important to note that the conductance peak values associated with the resonant states do not attain its maximum. The presence of evanescent states and energy level changes due to the exchange coupling may be some of the reasons. Another important contribution comes from the angular dependence of the transmission, as illustrated in Fig. \ref{fig.2b_G} (b) and (c). Partial maximum conductance values at particular $U_w$ denoted by dashed lines in panel (a) do exhibit maximum transmission but restricted to a finite angular range as explicitly shown in panels (b) and (c), leading to demoted conductance lesser than one quantum conductance. In fact, we can also observe that the transmission probability takes high values for angles around the normal incidence of the carriers. The sequence of conductance peaks allows a rich variety of spin and valley polarization dependence on the well depth, as shown in fig. \ref{fig.2b_G}(b), that may be explored.

The band alignments of the leads and well, and the transmission for the spin up bands $K_{+}$ and $K_{-}$ bands, shown in Fig. \ref{fig.6}, help in understanding the resonant features of the conductance and polarization results. The transmission for each band is shifted due to the presence of the EuS substrate. The transmission maxima are highlighted with dot symbols in the well bands, as depicted in  Fig. \ref{fig.6}(a) and (b).
 For normal incidence, as the gate potential is applied in the central region, the resonant levels (dots) begin to align with the Fermi energy at the leads (market with a black horizontal line) and, as a consequence, the conductance exhibits a series of sharp and well-defined peaks.  As expected, the number and energy distribution of these peaks depend on the well region length, which is a favorable condition to obtain gate-tunable values of spin and valley polarization.
 It is important to emphasize that the perfect resonant conductance peaks observed in Fig. 6 are obtained only for normal incidence, otherwise, an angle integrated conductance has to be calculated [via Eq. 9], giving broad peaks centered at the resonant energy values. These resonances can be tuned by the gate potentials, generating the corresponding spin- and valley-transport polarizations.

\begin{figure}[ht]
\centering
\includegraphics[width=1.0\linewidth]{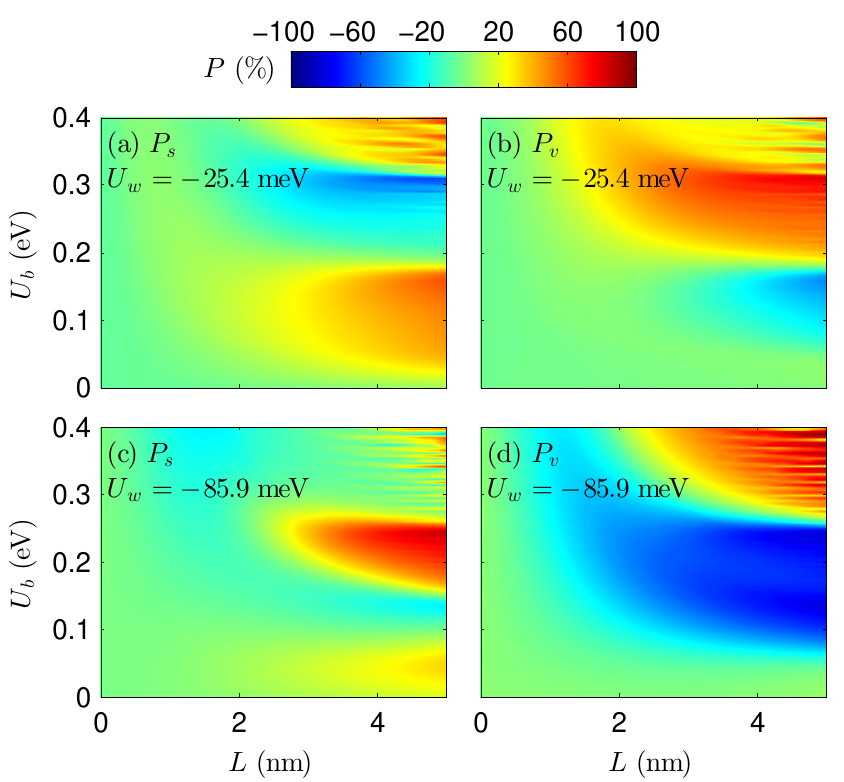}
\caption{Spin and valley polarization contour plots as functions of the barrier height $U_b$ and lengths $L$ for the double barrier system at a fixed energy equal to $104.68$ meV above the Fermi level. Well length is $d=30$ nm, well depth for panels(a-b) is $U_w=-25.40$ meV while for panels(c-d) is $U_w=-85.90$ meV.}
\label{fig.2b_contour_GP_Ub_L}
\end{figure}

Two relevant situations, for the spin and valley polarization showed in  Fig. \ref{fig.2b_G2} are highlighted with dashed lines in this plot: (i) a potential at which both spin and valley polarizations are zero ($-85.9$ meV) and (ii) a potential at which the valley and spin polarizations have opposite signs ($-25.4$ meV). We investigate how these polarizations depend on the barrier length $L$ and on the potential height $U_b$, at both energies, as depicted in Fig. \ref{fig.2b_contour_GP_Ub_L} (a-d). In the contour plots $U_w=-25.4$ meV and $-85.9$ meV for top and  bottom panels, respectively.
The results suggest that it is possible to obtain zero polarization or highly spin and valley polarized heterostructure depending on the potential well energy, in a wide range of parameter space ($U_b, L$). 
It is also observed that, by increasing the barrier potential intensity $U_b$ it is possible to change the spin and valley polarization signs, from positive to negative and vice-versa, at a fixed barrier length, specially around $L=4$ nm [see \ref{fig.2b_contour_GP_Ub_L} (a) and (d)]. As the barrier potential $U_b$ is increased, the energy distribution of the resonant states in the well region changes, separating and defining these levels in such a way that the resonant tunneling is affected. For some  $U_b$ values, the system conduces preferentially by the $\uparrow K_+$ or the $\uparrow K_-$ conductance components promoting, therefore, a  sign reversal of the spin and valley polarizations. Finally, for $d=30$ nm, the optimal parameter space values ($U_b, L$) at which the maximum valley polarization is obtained, is around $U_b=200$ meV and for $L>2.5$ nm.

\begin{figure}[ht]
\centering
\includegraphics[width=1.0\linewidth]{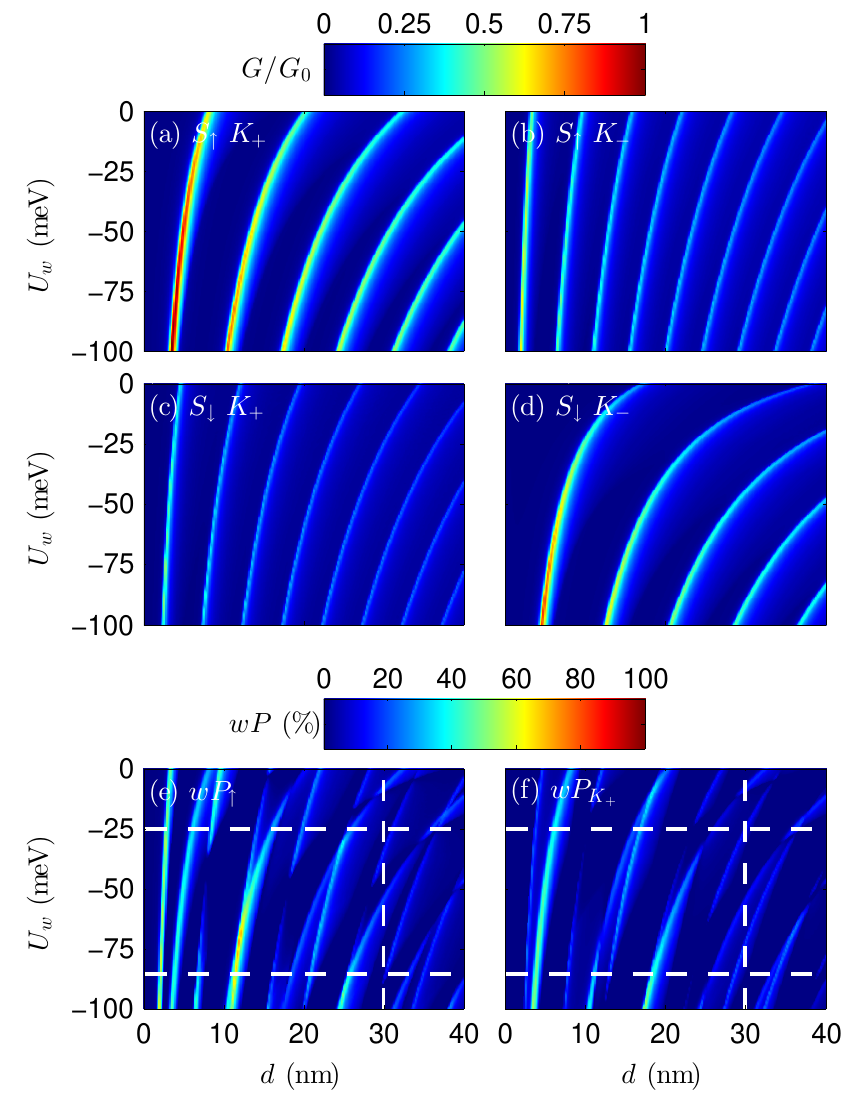}
\caption{Spin and valley conductance (a-d) and weighted polarization contour plots (e-f) as functions of the well depth and length, $U_w$ and $d$, for the double barrier system at a fixed energy of $104.68$ meV above the Fermi level. Barrier length and height are $L=3$ nm and $U_b=300$ meV, respectively.}
\label{fig.2b_contour_GP_Uw_d}
\end{figure}

We further explore the dependence of the valley and spin conductances and polarizations on the quantum well potential $U_w$ and length $d$. Results for each one of the conductance components, for a double barrier system of length and height $L=3$ nm and $U_b=300$ meV, respectively, and at a fixed incident electron energy $E=104.68$ meV above the Fermi level, are presented in Fig. \ref{fig.2b_contour_GP_Uw_d} (a-d). The resonant nature of the electronic states of the well are revealed through the parabolic-like features, marking non-zero conductance in the contour plots, as the intensity potential and geometrical dimension of the quantum well are swept.
The spin and valley polarization contour plots, depicted in Fig. \ref{fig.2b_contour_GP_Uw_d} (e) and (f), respectively, present complex patterns, indicating  clearly that drastic changes may occur for a fixed quantum well length as the well potential goes from zero to $-100$ meV. 
It is important to notice that, to avoid spurious results, we have calculated in these plots the weighted polarization \cite{t6,orellana2013spin}, which is defined as $wP_{_S} = G_{_{\uparrow(\downarrow)}}P_{_{S}}$ and $wP_{\nu}	= G_{_{K(')}}P_{{\nu}}$. 

\begin{figure}[ht]
\centering
\includegraphics[width=1.0\linewidth]{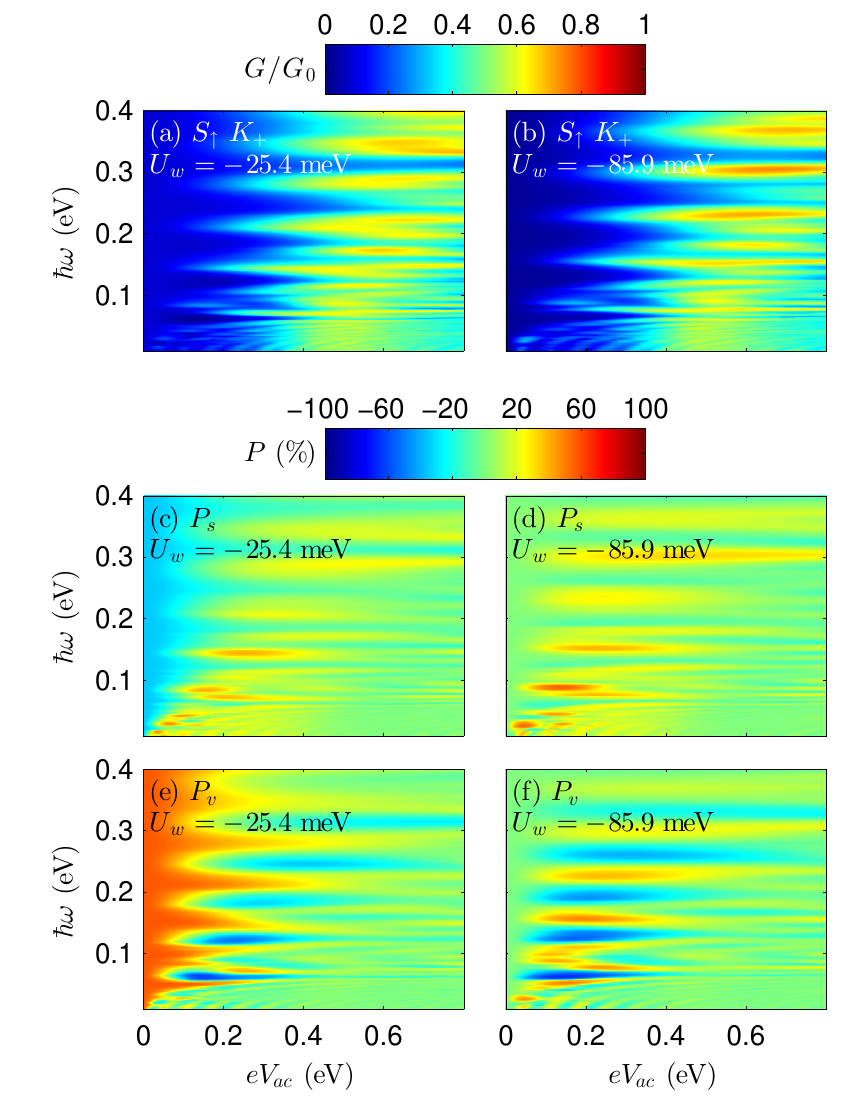}
\caption{Spin and valley conductance and polarization contour plots as functions of the ac-field frequency $\hbar\omega$ and power $eV_\text{ac}$, for the double barrier system in the FAR-IR frequency range. Parameters: barrier and well lengths $L = 3$ nm and $d=30$ nm, barrier height $U_b = 300$ meV. Well depth for the left (a,c,e) and right (b,d,f) panels, are $U_w=-25.4$ meV and $U_w=-85.9$ meV, respectively.}
\label{fig.2b_contour_GP_laser}
\end{figure}

\subsection{External AC-field}
As previously mentioned, we address the possibilities of getting interesting spin and valley filter scenarios for WSe$_2$ devices by properly exposing the system to a time-dependent radiation. The external potential can be a laser or a time dependent gate voltage, with frequency $\hbar \omega$ and amplitude potential $eV_{ac}$, applied to the whole system. In order to obtain modulations of the heterostructure transport response, we have explored different time dependent potential parameters ($\hbar \omega$, $eV_{ac}$). The main features obtained for a double barrier device, under the oscillating potential, are illustrated in Fig. \ref{fig.2b_contour_GP_laser} via spin and valley conductance [(a) and (b)] and polarization [(c)-(f)] contour-plots in the mid-infrared (between $413$ - $24.8$ meV) and far-infrared or Terahertz (between $24.84$ - $1.24$ meV) frequency range, and as a function of the amplitude potential $eV_{ac}$. The analyzed structure has a barrier height $U_b = 300$ meV and length $L= 3$ nm, whereas the quantum well geometry is defined by a well length $d=30$ nm and two potential depths $U_w = -25.40$ meV and $-85.90$ meV, in the left and right panels, respectively. These potential values are marked with dashed white lines in weighted  spin polarization  $\omega P_\uparrow$ and valley polarization $\omega P_{K+}$ of Fig. \ref{fig.2b_contour_GP_Uw_d} (e) and (f).
\begin{figure}[ht]
\centering
\includegraphics[width=1.0\linewidth]{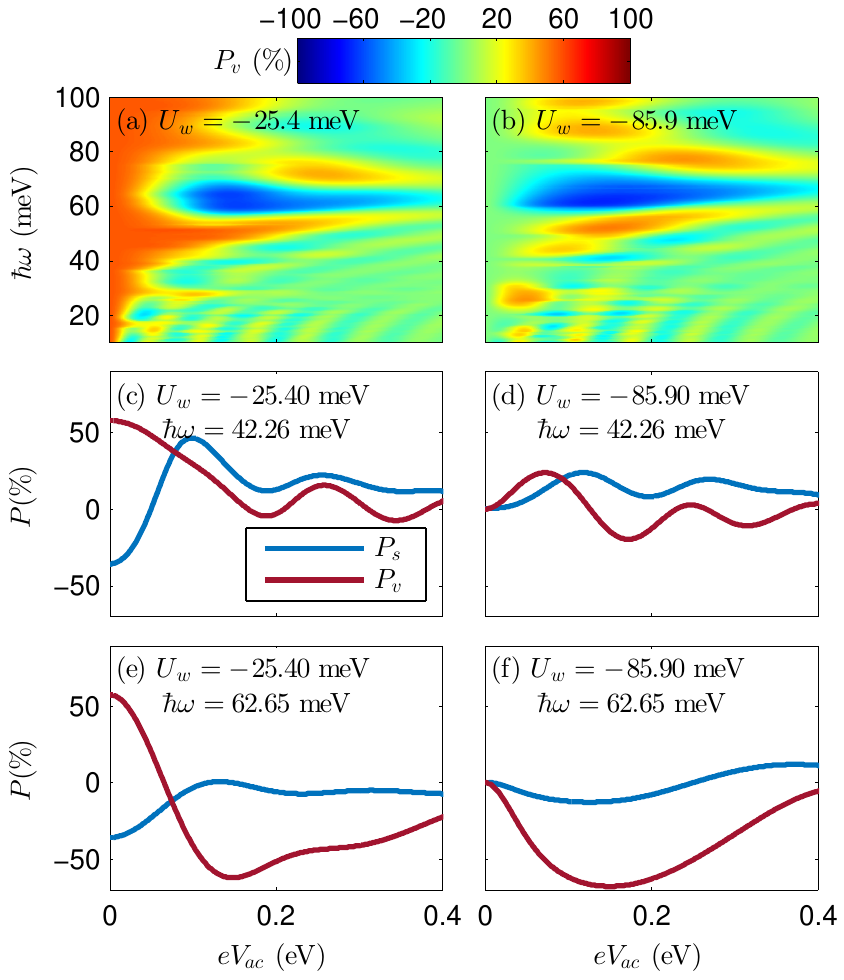}
\caption{(a-b) Zoom of the valley polarization contour plots mapped  in Fig. \ref{fig.2b_contour_GP_laser}(e-f) for $U_w$= -25.4 meV and -85.9 meV, respectively. (c-f) Valley and spin polarization results as a function of the ac-field power for the same double barrier system described in the previous figure: barrier length $L = 3$ nm and height $U_b = 300$ meV, well length $d = 30$ nm. The well depth are $U_w = -25.40$ meV (left panels) and $-85.90$ meV (right panels).  In all panels $E_{_F} = 849.98$ meV and an ac-field frequency $\hbar\omega =$ 42.26 meV and 62.64 meV, as marked in each panel.}
\label{fig.2b_P}
\end{figure}

The spin and valley resolved conductances (Fig. \ref{fig.2b_contour_GP_laser} (a) and (b)) show that, by applying an oscillating potential to the system, with a radiation amplitude $0< eV_{ac} < 0.8$ eV, both conductance components become different to zero, with an oscillating behavior as the frequency increase. This means that, due to the presence of the time dependent ac-field, the effective modulation of the electronic wavefunction quantum phase promotes the resonant tunneling in the double barrier heterostructure. Actually, variations up to $70\%$ are noticed at particular potential amplitudes and frequencies.

Similar features are observed in the valley and spin polarization maps, considering the same two values of the potential well $U_w$, as depicted in the contour plots of Fig. \ref{fig.2b_contour_GP_laser} (c-f).
For $U_w= - 25.4$ meV and $U_w= - 85.9$ meV, the spin polarization shows smooth and periodic modulations of its maximum value as the ac-field frequency is increased. However, there are narrow frequency ranges [orange regions in panels (c) and (d)] at which high spin polarization are observed, for different ac-field amplitude. This is more evident for a deeper well potential ($U_w= - 85.9$ meV) when more resonant states are allowed into the conductor region, and in the terahertz frequency range, up to 144 meV.
For the valley polarization, at $U_w= - 25.4$ meV, as the ac-field amplitude is turned on and increased, it is possible to switch from $K_+$ to $K_-$ valley filter, revealed by the sequence of orange and blue color in the contour plots of panel (e), for a fixed ac-field frequency. Also, considering the potential well $U_w= - 85.9$ meV, the system is moved from zero-spin and valley polarizations to $\pm 20\%$ and $\pm 60\%$ spin and valley polarizations, respectively, depending on the far-IR radiation frequency. 

These valley switch polarization features are highlighted in the zoom presented in Fig. \ref{fig.2b_P} (a-b), where the laser frequency and potential amplitude were constraint to smaller ranges. Some two-dimensional cuts for the valley and spin polarizations as a function of the ac-field  amplitudes are presented in Fig. \ref{fig.2b_P} (c-f), for fixed quantum well potential and frequency values, as its is indicated in the panels. Actually, by adjusting the ac-field potential, a variety of filter regimes may be achieved, which transforms the double quantum well geometry as a promising platform to reveal the spin and valley filtering behavior of WSe$_2$ monolayers.

\section{Summary}

In this work, we have investigated the spin and valley transport properties of a WSe$_2$ monolayer placed on top of a ferromagnetic insulator. Single and double barrier heterostructures were explored. 
By applying external potentials to the system, we have shown that the systems can be used as efficient valley and spin filter devices
and how the polarized transport properties depend on the gate-potential intensities and geometrical configuration.
We believe that the combination of heterostructured TMD geometries and appropriated gate potentials, proposed in our work, allows to control the resonant tuning provided by the state alignment with the electronic carrier bands. We have found that double barrier structures are more appropriated to valley and spin filters, due to the resonant tunneling, compared with single barrier systems.
Additionally, we investigate how the spin and valley polarizations are modified when an ac-field is applied to the system. 
The radiation field allows tuning both polarizations in a wide range of device geometries, radiation intensities and frequencies, especially in the terahertz range. Inversion of the spin and valley polarizations signal are found possible by changing the laser frequency for fixed power amplitudes. The possibility of spin and valley polarization inversion was not addressed in other works and may be considered as an advantage proposal.
Our results suggest that 2D WSe$_2$ heterostructures are good candidates to provide spin and valley dependent transport and can drive experimental efforts in order to probe spin and valley polarized currents.

\section{ACKNOWLEDGMENTS}
This work was partially financed by Fondecyt, Grants 1180914 and 1201876 of Chile. AL would like to thanks partial support from FAPERJ (grant E-26/202.567/2019), CNPq, and INCT de Nanomateriais de Carbono.

\appendix

\section{}

A Hamiltonian like the one described by Eqs. (\ref{H}, \ref{H0}, \ref{Hex}, \ref{Hg}) can always be written in the following form
\begin{equation}
	H = 
	\begin{pmatrix}
		\Delta_c     &  \hbar v k_- \\
		\hbar v k_+  &  \Delta_v    \\
	\end{pmatrix}	
	\text{.}
	\label{general_H}
\end{equation}
In our model $\Delta_c = S_z(\eta\lambda_c+B_c) + U(x) + m/2$, $\Delta_v = S_z(\eta\lambda_v-B_v) + U(x) - m/2$, $k_\pm = \eta k_x \pm ik_y$, $v$ is the Fermi velocity and $\eta=\pm 1$ is the valley index. Note that $\hbar^2v^2k_+k_- = \left( \hbar vk \right)^2 = c_k^2$.

A straight forward diagonalization of the low energy Hamiltonian defined by eq. (\ref{general_H}) leads to the following spinor for an incident electron in the conduction band
\begin{equation}
	V^\eta_{S_z,\pm} = \frac{1}{D_k}
	\begin{pmatrix}
		\eta c_k e^{-i\eta\theta}  \\
		b_k  \\
	\end{pmatrix}
	\text{,}
\end{equation}
where $b_k = \sqrt{ \Delta_-^2  + c_k^2 }\,-\,\Delta_-$, $D_k = \sqrt{ c_k^2 + b_k^2 }$ and $2\Delta_- = \Delta_c - \Delta_v$.

Considering the invariance in the $y$ direction and the structure shown in Fig. \ref{fig.scheme_2b} the wave functions for the double barrier system are
\begin{eqnarray}
	\Psi_{I}(x,y) &=& \left[ \frac{e^{ik_x x}}{D_k}
	\begin{pmatrix}
		\eta c_k e^{-i\eta\theta}  \\
		b_k  \\
	\end{pmatrix}\right. \nonumber \\
	&+&\left. r_{\eta,S_z}\frac{e^{-ik_x x}}{D_k}
	\begin{pmatrix}
		\eta c_k e^{-i\eta\left( \pi - \theta \right)}  \\
		b_k  \\
	\end{pmatrix}
	\right]e^{ik_y y}\text{,}
\end{eqnarray}
\begin{eqnarray}
	\Psi_{II}(x,y) &=& \left[ A\frac{e^{iq_x x}}{D_q}
	\begin{pmatrix}
		\eta c_q e^{-i\eta\varphi}  \\
		b_q  \\
	\end{pmatrix}\right. \nonumber \\
	&+&\left. B\frac{e^{-iq_x x}}{D_q}
	\begin{pmatrix}
		\eta c_q e^{-i\eta\left( \pi - \varphi \right)}  \\
		b_q  \\
	\end{pmatrix}
	\right]e^{ik_y y}\text{,}
\end{eqnarray}
\begin{eqnarray}
	\Psi_{III}(x,y) &=& \left[ C \frac{e^{iw_x x}}{D_w}
	\begin{pmatrix}
		\eta c_w e^{-i\eta\phi}  \\
		b_w  \\
	\end{pmatrix}\right. \nonumber \\
	&+&\left. D\frac{e^{-iw_x x}}{D_w}
	\begin{pmatrix}
		\eta c_w e^{-i\eta\left( \pi - \phi \right)}  \\
		b_w  \\
	\end{pmatrix}
	\right]e^{ik_y y}\text{,}
\end{eqnarray}
\begin{eqnarray}
	\Psi_{IV}(x,y) &=& \left[ E\frac{e^{iq_x x}}{D_q}
	\begin{pmatrix}
		\eta c_q e^{-i\eta\varphi}  \\
		b_q  \\
	\end{pmatrix}\right. \nonumber \\
	&+&\left. F\frac{e^{-iq_x x}}{D_q}
	\begin{pmatrix}
		\eta c_q e^{-i\eta\left( \pi - \varphi \right)}  \\
		b_q  \\
	\end{pmatrix}
	\right]e^{ik_y y}\text{,}
\end{eqnarray}
\begin{eqnarray}
	\Psi_{V}(x,y) &=& t_{\eta,S_z} \frac{e^{ik_x x}}{D_k}
	\begin{pmatrix}
		\eta c_k e^{-i\eta\theta}  \\
		b_k  \\
	\end{pmatrix}
	e^{ik_y y}\text{,}
\end{eqnarray}
where the coefficients $D_q$, $b_q$, $D_w$ and $b_w$ are defined in the same way as $D_k$ and $b_k$, but with the corresponding parameters accordingly to Fig. \ref{fig.scheme_2b}.

The transmission probability is found  by matching the wave functions at the interfaces; at $x=0$ we have $\Psi_{I}(0) = \Psi_{II}(0)$ which is written into two equations
\begin{widetext}

\begin{eqnarray}
	\frac{\eta c_k e^{-i\eta\theta}}{D_k} + r_{\eta,S_z}\frac{\eta c_k e^{-i\eta\left( \pi - \theta \right)}}{D_k} &=& A\frac{\eta c_q e^{-i\eta\varphi}}{D_q} + B\frac{\eta c_q e^{-i\eta\left( \pi - \varphi \right)}}{D_q} \label{eq1}\\
	\frac{b_k}{D_k} + r_{\eta,S_z}\frac{b_k}{D_k} &=& A\frac{b_q}{D_q} + B\frac{b_q}{D_q} \label{eq2}\text{.}
\end{eqnarray}

At $x=L$ we have $\Psi_{II}(L) = \Psi_{III}(L)$, and similarly
\begin{eqnarray}
	A\frac{\eta c_q e^{-i\eta\varphi}}{D_q}e^{iq_x L} + B\frac{\eta c_q e^{-i\eta\left( \pi - \varphi \right)}}{D_q}e^{-iq_x L} &=& C\frac{\eta c_w e^{-i\eta\phi}}{D_w}e^{iw_x L} + D\frac{\eta c_w e^{-i\eta\left( \pi - \phi \right)}}{D_w}e^{-iw_x L} \label{eq3}\\
	A\frac{b_q}{D_q}e^{iq_x L} + B\frac{b_q}{D_q}e^{-iq_x L} &=& C\frac{b_w}{D_w}e^{iw_x L} + D\frac{b_w}{D_w}e^{-iw_x L} \label{eq4}\text{.}
\end{eqnarray}

At $x=L+d$, $\Psi_{III}(L+d) = \Psi_{IV}(L+d)$ and we obtain
\begin{eqnarray}
	C\frac{\eta c_w e^{-i\eta\phi}}{D_w}e^{iw_x (L+d)} + D\frac{\eta c_w e^{-i\eta\left( \pi - \phi \right)}}{D_w}e^{-iw_x (L+d)} &=& E\frac{\eta c_q e^{-i\eta\varphi}}{D_q}e^{iq_x (L+d)} + F\frac{\eta c_q e^{-i\eta\left( \pi - \varphi \right)}}{D_q}e^{-iq_x (L+d)} \label{eq5}\\
	C\frac{b_w}{D_w}e^{iw_x (L+d)} + D\frac{b_w}{D_w}e^{-iw_x (L+d)} &=& E\frac{b_q}{D_q}e^{iq_x (L+d)} + F\frac{b_q}{D_q}e^{-iq_x (L+d)} \label{eq6}\text{.}
\end{eqnarray}

And finally at $x=L+d+L$, $\Psi_{IV}(2L+d) = \Psi_{V}(2L+d)$ and we get
\begin{eqnarray}
	E\frac{\eta c_q e^{-i\eta\varphi}}{D_q}e^{iq_x (2L+d)} + F\frac{\eta c_q e^{-i\eta\left( \pi - \varphi \right)}}{D_q}e^{-iq_x (2L+d)} &=& t_{\eta,S_z}\frac{\eta c_k e^{-i\eta\theta}}{D_k}e^{ik_x (2L+d)} \label{eq7}\\
	E\frac{b_q}{D_q}e^{iq_x (2L+d)} + F\frac{b_q}{D_q}e^{-iq_x (2L+d)} &=& t_{\eta,S_z}\frac{b_k}{D_k}e^{ik_x (2L+d)} \label{eq8}\text{.}
\end{eqnarray}

\end{widetext}

The $r_{\eta,S_z}^\text{Double}$ and $t_{\eta,S_z}^\text{Double}$ coefficients are obtained by solving the equation system defined by Eq. \ref{eq1} to Eq. \ref{eq8}, resulting in the following expression for the transmission coefficient:
\begin{equation}
	t_{\eta,S_z}^\text{Double} = \frac{ 16 e^{-2i k_x L} e^{-i k_x d} \cos\left(\varphi\right)^2 \cos\left(\phi\right) \cos\left(\theta\right) }{ \xi }\text{,}\label{t_double}
\end{equation}

with the denominator $\xi$ given by

\begin{eqnarray}
	\xi &=& F_1 + \left(\frac{c_k b_w}{c_w b_k} + \frac{c_w b_k}{c_k b_w}\right)F_2 + \left(\frac{c_k b_q}{c_q b_k} + \frac{c_q b_k}{c_k b_q}\right)F_3 + \nonumber\\
	    & &\left(\frac{c_q b_w}{c_w b_q} + \frac{c_w b_q}{c_q b_w}\right)F_4 + \left(\frac{c_k c_w b_q^2}{c_q^2 b_k b_w} + \frac{c_q^2 b_k b_w}{c_k c_w b_q^2}\right)F_5\text{.}\nonumber\\
\end{eqnarray}

where the functions $F_i$ are defined as

\begin{eqnarray}
	F_1 &=& - 8i   \sin\left(w_x d\right)   \cos\left(\theta + \phi\right)                          \nonumber\\
		& & + 8i   \sin\left(w_x d\right)   \cos\left(\theta - \phi\right)                          \nonumber\\
		& & +  4   e^{   i w_x d }   \cos\left(2 q_x L\right)   \cos\left(\theta + \phi\right)     \nonumber\\
		& & +  4   e^{ - i w_x d }   \cos\left(2 q_x L\right)   \cos\left(\theta - \phi\right)     \nonumber\\
		& & +  2   e^{ - i w_x d }   e^{ - 2i q_x L }   \cos\left(\theta + \phi + 2 \varphi\right) \nonumber\\
		& & +  2   e^{   i w_x d }   e^{ - 2i q_x L }   \cos\left(\theta - \phi + 2 \varphi\right) \nonumber\\
		& & +  2   e^{ - i w_x d }   e^{   2i q_x L }   \cos\left(\theta + \phi - 2 \varphi\right) \nonumber\\
		& & +  2   e^{   i w_x d }   e^{   2i q_x L }   \cos\left(\theta - \phi - 2 \varphi\right) \text{,}
\end{eqnarray}

\begin{eqnarray}
	F_2 &=& - 4i   \sin\left(w_xd\right)   \left( \cos\left(2 L q_x\right) + \cos\left(2 \varphi\right) \right) \text{,}
\end{eqnarray}

\begin{eqnarray}
	F_3 &=& - 4i \left( \sin\left(w_x d\right) + \sin\left(2 q_x L - w_x d\right) \right)   \cos\left(\varphi - \phi\right)  \nonumber\\
		& & + 4i \left( \sin\left(w_x d\right) - \sin\left(2 q_x L + w_x d\right) \right)   \cos\left(\varphi + \phi\right)  \text{,} \nonumber\\
\end{eqnarray}

\begin{eqnarray}
	F_4 &=& -  8   \sin\left(w_xd\right)   e^{   i q_x L }   \sin\left(q_x L\right)   \cos\left(\theta - \varphi\right)    \nonumber\\
		& & -  8   \sin\left(w_xd\right)   e^{ - i q_x L }   \sin\left(q_x L\right)   \cos\left(\theta + \varphi\right)   \text{,} \nonumber\\
\end{eqnarray}
and
\begin{eqnarray}
	F_5 &=& 8i   \sin\left(w_xd\right)   \sin\left(q_xL\right)^2 \text{.}
\end{eqnarray}

For the single barrier heterostructure the wave functions for the zones III and IV must be neglected and $d$ must be set equal to zero. A similar equation system is then obtained. The transmission coefficient of the single barrier,  $t_{\eta,S_z}^\text{Single}$,  is analytically determined, and given by Eqs. (\ref{tsingle1}) and (\ref{tsingle2}).

\vspace{1cm}
~

~


\end{document}